\documentclass[12pt,a4paper,psamsfonts]{article}
\usepackage{amsmath,amssymb,amsthm}
\usepackage{eucal}
\usepackage{geometry}

\numberwithin{equation}{section}

\def\a{\alpha}
\def\b{\beta}
\def\g{\gamma}
\def\d{\delta}

\def\k{\kappa}
\def\l{\lambda}
\def\m{\mu}
\def\n{\nu}

\def\r{\rho}

\def\s{\sigma}

\def\ome{\omega}

\def\G{\Gamma}

\def\L{\Lambda}

\def\RR{\mathbb R}

\newcommand{\C}[1]{$(\ref{#1})$}
\newcommand{\hph}[1]{{\hphantom{#1}}}
\def\o{\over}
\def\pa{\partial}

\def\hlf{\frac{1}{2}}
\def\lp{\left(}
\def\rp{\right)}

\makeatletter
\renewcommand\section{\@startsection {section}{1}{\z@}%
                                   {-3.5ex \@plus -1ex \@minus -.2ex}
                                   {2.3ex \@plus.2ex}%
                                   {\normalfont\large\bfseries}}
\renewcommand\subsection{\@startsection{subsection}{2}{\z@}%
                                     {-3.25ex\@plus -1ex \@minus -.2ex}%
                                     {1.5ex \@plus .2ex}%
                                     {\normalfont\bfseries}}
\makeatother

\begin{document}

\begin{center}
\addtolength{\baselineskip}{.5mm}
\thispagestyle{empty}
\begin{flushright}
{\sc MIFPA-14-38}\\
\end{flushright}

\vspace{20mm}

{\Large  \bf Kaluza-Klein Theories Without Truncation}
\\[15mm]
{Katrin Becker, Melanie Becker and Daniel Robbins}
\\[5mm]
{\it George P. and Cynthia W. Mitchell Institute for }\\
{\it Fundamental Physics and Astronomy, Texas A\& M University,}\\
{\it College Station, TX 77843-4242, USA}\\[5mm]

\vspace{20mm}

{\bf  Abstract}
\end{center}

In this note we will present a closed expression for the space-time effective action for all bosonic fields (massless and massive) obtained from the compactification of gravity or supergravity theories (such as type II or eleven-dimensional supergravities) from $D$ to $d$ space-time dimensions.

\vfill
\newpage

\tableofcontents

\section{Introduction}

A diversity of matter fields and interactions in a $d$-dimensional space-time can sometimes be described
by a simpler theory in higher, say $D$, dimensions. This idea was pioneered almost 100 years ago by Kaluza and Klein in an attempt to unify four-dimensional electromagnetism and gravity in terms of five-dimensional gravity. It is still one of the most intriguing approaches to unify gravitation with other forces in nature. It unifies different phenomena in $d$ dimensions and makes predictions. In the original Kaluza and Klein case one massless scalar was predicted.

Supergravity theories in $D=10$ or $D=11$ dimensions are a natural starting point.
In this note we take the ground state to be a direct product of $d$-dimensional
Minkowski space-time, $M^d$, and a smooth and compact Riemannian manifold $Y$.
We are primarily interested in preserving supersymmetry in space-time which requires the holonomy group of the metric on $Y$ to be a certain
subgroup of the orthogonal group. The different possibilities are on Berger's list.

We then wish to describe the fluctuations about the ground state. Our goal is to construct the $d$-dimensional space-time effective action for all fields, not only the massless sector.
To do this
our guiding principles are locality in $d$ and $D-d$ dimensions. Locality in $d$
dimensions is achieved by letting the fields depend covariantly
on space-time coordinates. Moreover, these fields
are also functions or forms along $Y$. This is locality in $D-d$ dimensions.
The coordinates on the internal space are then interpreted as labels
in the $d$-dimensional theory. The results are most naturally written using abbreviated ``DeWitt'' notation \cite{dewitt}.

In this note we will present a closed expression for the space-time effective
action for all fields (massless and massive) obtained from the compactification of type II and eleven-dimensional supergravities to $d$ space-time dimensions. In a forthcoming publication \cite{progress} we will present the manifestly supersymmetric actions. Some of the interactions were predicted in ref.\ \cite{Becker:2014rea}.  With the component actions at hand it should be possible to obtain the full action in superspace and compare with the predictions of ref.\ \cite{Becker:2014rea}.  Our approach has similarities with the program begun in refs. \cite{deWit:1986mz}, in which eleven-dimensional supergravity fields were written according to a $4+7$ split of the space-time coordinates, as an intermediate step to making manifest a local $\operatorname{SU}(8)$ symmetry.  However, the goal of the present work is to give a completely general result for the reduction of gravitational theories.

In section 2 we study the compactification of the Einstein-Hilbert action to arbitrary space-time dimensions. We present a closed expression for the space-time effective action for fields arising
from the metric and we analyze the non-abelian gauge symmetry arising from diffeomorphisms. In section 3  antisymmetric tensor fields are discussed.
We consider both kinetic and Chern-Simons terms. In the summary we present
the complete action for all fields in an example. Using the results presented in this paper an action for all fields for the compactification of any supergravity theory to any number of space-time dimensions can be written down.

\section{Einstein-Hilbert action}

The Einstein-Hilbert action in $D$ space-time dimensions is\footnote{\label{indcon} In what follows, indices
$M,N,\dots$, $\m,\n,\dots$ and $a,b,\dots$ are tangent to the $D$-, $d$-
and $D-d$-dimensional spaces, respectively.  We always sum over repeated indices}
\begin{equation}\label{eds}
S = {1\o 2 \k^2}\int d^{D} x \sqrt{-G} q  R.
\end{equation}
Here we have included an unspecified function $q$. This function can depend on fields (for example, the dilaton
in type II supergravity in the string frame) but does not depend on the metric. In $D \neq 2$ dimensions $q$
can be removed by a metric rescaling, but often we may wish to work in a frame (like the string frame) which includes non-trivial $q$.

We wish to construct the action for the
fluctuations about the ground state with metric $\hat G_{MN}$. To do this
we expand the metric $G_{MN}$ about this ground state
\begin{equation} \label{ai}
G_{MN} = \hat G_{MN} + \d G_{MN}.
\end{equation}
The Christoffel symbol becomes
\begin{equation} \label{aii}
\G_{MN}^L = \hat \G_{MN}^L + \d \G_{MN}^L,
\end{equation}
where
\begin{equation}\label{aiii}
\d \G_{MN}^L ={1\o 2} G^{LR} \left(\hat \nabla_M G_{RN}  + \hat \nabla_N G_{RM} -\hat \nabla_R G_{MN}  \right),
\end{equation}
is a tensor.
After partial integration the gravitational action becomes
\begin{equation}\label{main}
\begin{split}
S =  {1\o 2 \k^2} \int d^D x \sqrt{-G}  q \Big[&
G^{MK}\hat R_{MK} +
2G^{M[N} G^{P]Q} \hat \nabla_M (\log q) \hat \nabla_{ N}  G _{PQ}   \\
+ &
 \left(G^{PU} G^{Q[T} G^{R]S} - {1\o 2} G^{RU} G^{P[S} G^{Q]T} \right)\hat \nabla_R G_{PQ} \hat \nabla_U G_{ST}.
\Big].
\end{split}
\end{equation}
Note that this result includes all orders in the perturbations of the
metric. The inverse metric $G^{MN}$ will, in general, be an infinite expansion since it
is obtained by inverting eqn.\ \C{ai}.
Here and in the following hatted quantities refer to the ground state.
So, for example,
$\hat \nabla$ is the Levi-Civita connection of the background metric.

\subsection{Compactification}

Next we take the ground state to be $M^d \times Y$, where
$M^d$ is $d$-dimensional Minkowski
space-time with coordinates $x^\m$, $\m=0,\dots,d-1$ and with metric $\hat G_{\m\n} = \eta_{\m\n}$
and $Y$ is a $(D-d)$-dimensional internal manifold with
coordinates $y^a$, $a=d,\dots,D$ and
metric $\hat G_{ab}=\hat g_{ab}(y)$. We take the holonomy group of the metric on $Y$ to be a subgroup of the orthogonal group, which leads to some amount of unbroken supersymmetry in space-time.
Moreover, we remark here
that we are not interested in engineering a ground state with particular properties. Rather we take the internal space to be the most generic special holonomy manifold and we wish to describe the fluctuations about this ground state in full generality.

The fluctuations are encoded in the metric
\begin{equation} \label{bi1}
G_{MN} = \begin{pmatrix}
h _{\m\n} + g _{cd} A^c_\m A^d_\n && g _{bc} A^c_\m \\
g _{ac} A^c_\n && g _{ab}
\end{pmatrix},
\end{equation}
and its inverse
\begin{equation} \label{bi2}
G^{MN} = \begin{pmatrix}
h^{\m\n} & - h^{\m \r} A^b _\r \\
-h^{\n\r} A^a_\r &g^{ab}+ h^{\r\s} A^a_\r A^b_\s
\end{pmatrix} .
\end{equation}
Here
\begin{equation}
h _{\m\n} = h_{\m \n} (x,y), \quad g _{ab} = g _{ab}(x,y), \quad A_\m^a = A_\m ^a(x,y).
\end{equation}
The fields depend on $x^\m$, since these are $d$-dimensional fields, while
the $y^a$ dependence is interpreted as a continuous label carried
by the space-time fields.  Any $D$-dimensional metric can be written in this form. Note that the inverse metric contains terms which are at most quadratic in $A_\m^a$, and as we will see next the action contains terms which are at most quartic. The ground state corresponds to
\begin{equation}
h_{\m\n} = \eta_{\m\n}, \qquad g_{ab} = \hat g_{ab}(y), \qquad A^a_\m=0.
\end{equation}

\subsection{The action}

Using the metric \C{bi1} and its inverse \C{bi2} the action \C{main} becomes
\begin{equation}
S=S_{\rm kin}+S_{\rm pot} + S_{\rm gauge},
\end{equation}
with kinetic terms
\begin{equation}
\begin{split}
S_{\rm kin}=  {1\o 2 \k^2}
 \int dV \Big[&
\big( h ^{ \b \m }h ^{\g[ \r} h ^{\a] \n} - {1\o 2} h ^{\a\m} h ^{\b[\n} h ^{\g] \r}\big)
{\cal D}_\a h _{\b\g} {\cal D}_{\m} h _{\n \r }\\
&- h^{\m[\n}h^{\s]\r}g^{ab}{\cal D}_\m h_{\r\n}{\cal D}_{\s}g_{ab}
+  \hlf h ^{\m\n}g^{a[b}g^{c]d} {\cal D}_\m g _{ab}{\cal D}_{\n}g_{cd}   \\
&+ 2 h^{\m[ \n}h^{\rho] \s} {\cal D}_\m (\log{q}){\cal D}_{ \n} h_{\rho \s}  +
h^{\m\n}g^{ab} {\cal D}_\m (\log{q} ){\cal D}_\n g _{ab}\Big],
\end{split}
\end{equation}
potential terms
\begin{equation}
\begin{split}
S_{\rm pot} =  {1\o 2 \k^2}
 \int dV\Big[&
\hlf h^{\m[\n}h^{\s]\r}g^{ab}\hat\nabla_a h_{\m\n}\hat\nabla_{b}
h_{\r\s}
+ h ^{\m\n}g^{a[b}g^{c]d}
\hat\nabla_a h _{\m\n}\hat \nabla_b g_{cd} \\
& + g^{ab}g^{cd}g^{ef}\big( -\hat \nabla_a g _{c [b }\hat \nabla_{|d|}  g _{e] f}+
\hlf\hat \nabla_a g _{c[ d}
\hat \nabla_{|b|}  g _{e]f} \big)\\
& +
h^{\m\n}g^{ab} \hat \nabla_a({\log q})\hat \nabla_b h_{\m\n} +2 g^{a[ b}g^{c] d} \hat \nabla_a (\log q)\hat \nabla_{ b} g _{c d} \Big],
\end{split}
\end{equation}
and gauge field action
\begin{equation}
\begin{split}
S_{\rm gauge}= & - {1\o 8 \k^2}
\int d V h^{\m\n}h^{\rho\s}g _{ab}{\cal F}^a_{\m\rho}{\cal F}^b_{\n\s}.
\end{split}
\end{equation}
Here
\begin{equation}
\label{eq:dV}
d V  = d^dxd^{D-d}y\, q \sqrt{-h}\sqrt{g},
\end{equation}
with $\hat g= \det \hat g_{ab}(y)$ and $g = \det g_{ab}(x,y)$. Moreover, we have defined the field strength
\begin{equation}\label{gfield}
{\cal F }^a_{\m\n} = 2\pa _{[\m}A^a_{\n]}-2A^b_{[\m}\hat \nabla_{|b|}A^a_{\n]},
\end{equation}
and covariant derivatives
\begin{equation}\label{covs}
\begin{split}
{\cal D}_\m q &= \pa _\m q-\hat \nabla_aqA^a_\m,\\
{\cal D }_\m g_{ab} &= \pa _\m g_{ab}-g_{ac}\hat \nabla_bA^c_\m-g_{bc}\hat \nabla_aA^c_\m-
\hat \nabla_c g_{ab}A^c_\m, \\
{\cal D }_\m h_{\n\rho} &= \pa _\m h_{\n\rho}-\hat \nabla_a h_{\n\rho}A^a_\m.
\end{split}
\end{equation}
In section 2.2 we will explain the symmetry of the space-time effective action arising from
$D$-dimensional covariance. This will determine the choice of field strength and covariant derivatives.

Since the ground state metric is Ricci flat we have set
\begin{equation}
\hat R_{\m\n}=0, \qquad \hat R_{ab}=0.
\end{equation}
But note that in deriving the space-time effective action we have not used the Ricci flatness of the internal
space anywhere.
A compactification on a non-Ricci flat space, say a sphere, leads to the effective action presented
in section 2.2 together with an additional term obtained by taking the first term in the bracket of eqn.
\C{main} into account.

This action can be further rewritten to obtain canonical kinetic terms, for example. While
the above result is general, further simplifications are case dependent. Both $h_{\m\n}$ and $g_{ab}$ can
be further rescaled to obtain an action with some desired properties.
For example, as we show below in
the context of eleven-dimensional supergravity,
$h_{\m\n}$ can be rescaled
by appropriate functions to obtain canonical kinetic terms. It is straightforward but tedious to show
that canonical kinetic terms can indeed be obtained for any value of $D$ and for $d>2$.

\subsection{Symmetries}

In general relativity the space-time is a Riemannian manifold. The physical equations are covariant in the sense that they preserve their form under coordinate transformations $x^M \to x'^M$. If the coordinate transformation is close to the identity we set
\begin{equation}
x^M \to x'^M = x^M - \xi^M(x),
\end{equation}
where $\xi^M$ is a vector field.
The metric then changes by the infinitessimal amount
\begin{equation}\label{metrictrafo}
\d G_{MN}(x) = G'_{MN}(x)-G_{MN}(x) =  \xi^R\pa_R G_{MN} +G_{RN}\pa_M \xi^R +G_{MR}\pa_N \xi^R .
\end{equation}
Once compactified, covariance in $D$ dimensions
gives rise to the symmetries of the space-time effective theory in $d$ dimensions.

Next we derive how space-time fields change after coordinate transformations with parameters
$\xi^a=\xi^a(x,y)$ and $\xi^\mu= \xi^\m(x,y)$.
Lets consider the coordinate transformations with parameter $\xi^a$ first.
Given the transformation of the metric in eqn.\C{metrictrafo} it is a small exercise to determine how
space-time fields transform. The result is
\begin{equation}\label{gaugetrafo}
\begin{split}
\d q &= \hat \nabla_aq\xi^a,\\
\d g_{ab} &= \hat \nabla_cg_{ab}\xi^c+g_{ac}\hat \nabla_b\xi^c+g_{bc}\hat \nabla_a\xi^c, \\
\d h_{\m\n} &= \hat \nabla_a h_{\m\n}\xi^a,\\
\d A^a_\m &= \pa _\m\xi^a+\hat \nabla_bA^a_\m\xi^b-A^b_\m\hat \nabla_b\xi^a.
\end{split}
\end{equation}
The covariant derivatives defined in eqn.\ \C{covs}
transform nicely (no space-time derivatives of the
gauge parameter appearing) under coordinate transformations with parameter $\xi^a$. Indeed,
\begin{equation}
\begin{split}
\d\lp{\cal D }_\m q\rp &= \hat \nabla_a\lp{\cal D }_\m q\rp\xi^a,\\
\d\lp{\cal D }_\m g_{ab}\rp &= \hat \nabla_c\lp{\cal D }_\m g_{ab}\rp\xi^c+{\cal D }_\m g_{ac}\hat \nabla_b\xi^c+{\cal D }_\m g_{bc}\hat \nabla_a\xi^c, \\
\d\lp{\cal D }_\m h_{\n\rho}\rp &= \hat \nabla_a\lp{\cal D }_\m h_{\n\rho}\rp\xi^a,
\end{split}
\end{equation}
while for the field strength we find
\begin{equation}
\d{\cal F }^a_{\m\n} = \hat \nabla_b{\cal F }^a_{\m\n}\xi^b-{\cal F }^b_{\m\n}\hat \nabla_b\xi^a.
\end{equation}

Next consider coordinate transformations with parameter $\xi^\m$. We find
\begin{equation}
\begin{split}
\d q &= \pa _\m q\xi^\m,\\
\d g_{ab} &= \pa _\m g_{ab}\xi^\m+g_{ac}A^c_\m\hat \nabla_b\xi^\m+g_{bc}A^c_\m\hat \nabla_a\xi^\m, \\
\d h_{\m\n} &= \pa _\rho h_{\m\n}\xi^\rho+ h_{\m\rho}\pa _\n\xi^\rho+ h_{\n\rho}\pa _\m\xi^\rho- h_{\m\rho}A^a_\n\hat \nabla_a\xi^\rho- h_{\n\rho}A^a_\m\hat \nabla_a\xi^\rho, \\
\d A^a_\m &=  h_{\m\n}g^{ \,ab}\hat \nabla_b\xi^\n-A^b_\m A^a_\n\hat \nabla_b\xi^\n+A^a_\n\pa _\m\xi^\n+\pa _\n A^a_\m\xi^\n.
\end{split}
\end{equation}
General coordinate transformations of the $d$-dimensional space-time correspond to transformations with parameter $\xi^\mu=\xi^\mu(x)$ with no $y$ dependence, and for $d$-dimensional Minkowski space the global Poincar\'e transformations correspond to $\xi^\mu=a^\mu+\Lambda^\mu_{\hphantom{\mu}\nu}x^\nu$. However, the above transformations are more general. We note that when combined with the non-abelian gauge transformations in space-time explained in detail in the next section the external diffeomorphisms generated by $\xi^\m$ give rise to an algebra which extends the Poincar\'e algebra in a non-trivial way. It will be fascinating to further study properties of the resulting algebras \cite{progress}.

\subsection{The non-abelian gauge symmetry in space-time}

The diffeomorphism group on the internal manifold $Y$ is the group of all one-to-one differentiable maps
of $Y$ onto itself. The inverse maps are also differentiable. The group multiplication is the composition of maps. This group is denoted by Diff($Y$). A diffeomorphism that is sufficiently close to the
identity can be interpreted as a coordinate transformation
\begin{equation}
y^a \to y'^a = y^a-\xi^a,
\end{equation}
for some vector field $\xi^a=\xi^a(x,y)$. We take the $x$ dependence to be arbitrary but fixed.
As we explain next
Diff($Y$) appears as a gauge symmetry in space-time. It is an unconventional gauge group, since
it is infinite dimensional. The structure constants of the associated Lie algebra can be found in ref.\ \cite{dewitt}, for example.

To interpret Diff($Y$) as gauge symmetry it is most convenient to use
abbreviated ``DeWitt'' notation \cite{dewitt}. We use this notation for all indices pertaining to the
internal space while we keep the space-time indices explicit.
From the $d$-dimensional space-time point of view, $y$ should be viewed as part of the field label rather than as a coordinate.  To make this manifest, we will write, for instance
\begin{equation}
A^a_\m(x) =A_\m^a(x,y),
\end{equation}
where we have suppressed the $y$ dependence and where $a$ now stands for the index combination $(a;y)$.
This combination
will be considered as a composite index labeling the $d$-dimensional gauge fields.  A sum over field labels should then include an integral over $y$ as well as a sum over $a$. A prime on an index, say $A^{a'}_\m(x)$, is a condensed notation for $A^a_\m(x,y')$.

Let us briefly review the situation for a finite dimensional gauge group with the
aim of generalizing to the infinite dimensional case. The reasoning below
will later also be applied to tensor fields.
In the finite dimensional case
the infinitesimal gauge transformations and the field strength are related to the structure constants by
\begin{equation}
\d A^i_\m=\pa_\m\l^i+f^i_{\hph{i}jk}A^j_\m\l^k,\qquad F^i_{\m\n}=2\pa_{[\m}A^i_{\n]}+f^i_{\hph{i}jk}A^j_{[\m} A^k_{\n]},
\end{equation}
where  $f^i_{\hph{i}jk}$ are the structure constants and
$i$, $j$, and $k$ run over a basis for the Lie algebra.
The transformations on $A^i_\m$ close if
the structure constants satisfy the familiar properties
\begin{equation}
f^i_{\hph{i}jk}+f^i_{\hph{i}kj}=0,\quad f^i_{\hph{i}[j|m|}f^m_{\hph{m}k\ell]}=0,
\end{equation}
and in this case we have
\begin{equation}
[\d_1,\d_2]A^i_\m=\d_3A^i_\m, \qquad {\rm with } \qquad
\l_3^i=f^i_{\hph{i}jk}\l_1^j\l_2^k.
\end{equation}

If the gauge transformations act linearly on the space of scalars $\phi^m$,
\begin{equation}
\d\phi^m=\lp t_i\rp^m_{\hph{m}n}\phi^n\l^i,
\end{equation}
then closure of the transformations on $\phi^a$, with the same commutation as above, requires the matrices $(t_i)^m_{\hph{m} n}$ to satisfy
\begin{equation} \label{eq:LieRep}
[t_i , t_j] = f^k_{\hph{k}ij} t_k.
\end{equation}
This is just the statement that the $\phi^m$ transform as a representation of the gauge group.

Turning to the case of interest we note that the variation of $A_\m^a$ in eqn.\ \C{gaugetrafo} can be
recast in the form
\begin{equation} \label{gaugetrafo2}
\begin{split}
\d A^a_\m= \pa _\m\l^a+f^a_{\hph{a}b' c'' }A^{b'}_\m\l^{c''}.
\end{split}
\end{equation}
where
\begin{equation} \label{eq:IntDiffStructConsts}
f^a_{\hph{a} b' c'' }=\d_b^a \pa _c\d(y-y')\d(y-y'')-\d_c^a\d(y-y') \pa _b\d(y-y''),
\end{equation}
are the structure constants of the diffeomorphism group.
Indeed, note that
in uncondensed notation the second term on the right hand side of eqn.\ \C{gaugetrafo2} is
\begin{equation}
\int d^{D-d}y'd^{D-d}y''\left[ \d_b^a \pa _c\d(y-y')\d(y-y'')-\d_c^a\d(y-y') \pa _b\d(y-y'')\right] A^b_\m(y')\xi^c(y''),
\end{equation}
which by explicitly evaluating the integral becomes
\begin{equation}
\nabla_b A^a_\m\xi^b-A^b_\m\nabla_b\xi^a.
\end{equation}
Here we have identified $\l^a=\xi^a$.  In the last step we used that the dependence on Christoffel symbols cancels out of this expression, and we have suppressed the $x$ dependence.
Moreover, in abbreviated notation the field strength \C{gfield} is
\begin{equation}
{{\mathcal{F}}}^a_{\m\n}=2 \pa _{[\m} A_{\n]}^{a}+f^{a}_{\hph{a}b'c'' }A_\m^{b'}A_\n^{c''},
\end{equation}
as can be easily verified.

The scalar fields, $q$ and $g_{ab}$, both transform linearly (and don't mix) under these gauge transformations.  As above, if we use abbreviated notation $q^y=q(x,y)$ and
$g_{ab}(x)=g_{ab}(x,y)$, we have the representations
\begin{equation} \label{tone}
\lp t_a\rp^{y'}_{\hph{y'} y'' }=\d(y-y') \pa _a\d(y-y''),
\end{equation}
and
\begin{equation} \label{ttwo}
\lp t_{a}\rp_{(bc)'}^{\hph{(bc)'}(de)''}=\d^{(d}_{(b}\d^{e)}_{c)}\d(y-y') \pa _a\d(y-y'')+2\d^{(d}_a\d^{e)}_{(b} \pa _{c)}\d(y-y')\d(y-y'').
\end{equation}
Again, these have been chosen to match the previous expressions for $\d q$ and $\d g_{ab}$.  And again, it is a short calculation to show that these do indeed furnish a representation of the infinite dimensional non-abelian group by verifying that eqns. \C{tone}
and \C{ttwo} satisfy eqn.\ \C{eq:LieRep}. We also note that the covariant derivatives defined
in eqn.\ \C{covs} take the form
\begin{equation}\label{covi}
{\cal D}_\m \phi^{ab\dots} = \pa_\m \phi^{ab\dots}-A_\m^{b'} (t_{b'})^{ab\dots}_{\hph{ab\dots} (ab\dots)''}\phi^{(ab\dots)''},
\end{equation}
for any field $\phi^{ab\dots} = \phi^{ab\dots}(x,y)$ transforming
in some representation of the gauge group. So, for example,
\begin{equation}
{\cal D}_\m q^y = \pa_\m q^y - A^{b'}_\mu (t_{b'})^y_{\hph{y} y''}q^{y''} ,
\end{equation}
with $t$ given in eqn.\ \C{tone}.

So far very few assumptions have been made about the internal space $Y$
and by keeping locality on $Y$ manifest we obtained quite general closed expressions.
However, if desired these expressions can be further transformed in a case dependent manner.
Say if $Y=S^1$ the fields
and parameters can be Fourier expanded if periodic boundary conditions are imposed. As can be
seen by Fourier transforming eqn.\ \C{eq:IntDiffStructConsts}, in this case
the gauge symmetry is the Virasoro
algebra without central extension as has been realized in ref.\ \cite{Dolan:1983aa}, for example.

\subsection{Eleven-dimensional supergravity}

In this subsection we elaborate the example of eleven-dimensional supergravity
\cite{Cremmer:1978km} compactified to $d$ space-time dimensions. Compared to the previous section we further rescale the space-time metric to obtain canonical kinetic terms.

The Einstein-Hilbert action in eleven dimensions is
\begin{equation}\label{eds}
S = {1\o 2 \k^2}\int d^{11} x \sqrt{-G}  R .
\end{equation}
We take the eleven-dimensional metric to be of Kaluza-Klein form
\begin{equation} \label{bi}
G_{MN} = \begin{pmatrix}
f h _{\m\n} + g _{cd} A^c_\m A^d_\n && g _{bc} A^c_\m \\
g _{ac} A^c_\n && g _{ab}
\end{pmatrix} .
\end{equation}
Here we have rescaled the fields $h_{\m\n}$ by the function
\begin{equation}
\label{eq:fDef}
f = \lp \hat g \o  g \rp^k  , \qquad k={1\o d-2},
\end{equation}
to obtain a canonically normalized Einstein-Hilbert action in $d>2$ dimensions.

The effective action separates into a kinetic piece
\begin{equation}
\begin{split}
S_{\rm kin} =  - &  {1\o 8 \k^2} \int dv h ^{\a\b} \left[
\left(k  g^{ab}g^{cd}+g^{ac}g^{bd}\right){\cal D}_\a g _{ab}{\cal D}_\b g_{cd}\right]   \\
+ &  {1\o 2 \k^2} \int d v \left(
h ^{ \b \m }h ^{\g[ \r} h ^{\a] \n} - {1\o 2} h ^{\a\m} h ^{\b[\n} h ^{\g] \r}\right)
{\cal D}_\a h _{\b\g} {\cal D}_{\m} h _{\n \r },
\end{split}
\end{equation}
potential terms
\begin{equation}
\begin{split}
S_{\rm pot} = &    {1\o 4 \k^2} \int d v\lp \hat g \o g \rp^k
 \Big\{ g^{ab}h^{\a[\b}h^{\m]\n}\hat  \nabla_a {h }_{\a\b}\hat \nabla_b {h }_{\m\n}
\\
& -h^{\a\b}\lp k  g^{ab}g^{cd}+g^{ac}g^{bd}\rp
\hat \nabla_a h_{\a\b}\hat\nabla_b g_{cd}
 \\ & +
\left[ g^{pt} g^{qu} g^{rs} -
{1\o 2} g^{ps} g^{qt} g^{ru} +
2 k   g^{pr} g^{qu} g^{st} + \frac{4-d}{2}
k  ^2 g^{pq} g^{ru} g^{st} \right]
\hat \nabla_r g _{pq} \hat \nabla_u g_{st}
\Big\},
\end{split}
\end{equation}
and the gauge field action
\begin{equation}
S_{\rm gauge} =  -   {1\o 8 \k^2} \int dv  \lp g \o \hat g \rp^k h^{\a\b}
{h }^{\m \n}  g _{ab} {\cal F}_{\a\m}^a {\cal F}_{\b\n}^b,
\end{equation}
where
\begin{equation}
dv =  d^d x d^{11-d} y \sqrt{-h}\sqrt{\hat g }.
\end{equation}
The above expressions are valid if $d \neq 2$. In two dimensions the Einstein-Hilbert action is
scale invariant and a canonically normalized Eintein-Hilbert action cannot be obtained. As discussed
in section 2.3 this action describes an unconventional gauge theory with gauge group Diff($Y$).

Again, our goal is generality, which we achieve by keeping locality along $Y$ manifest. If however $Y=S^1$, for example, the fields can be Fourier transformed. The above action then describes a finite set of massless fields and an infinite tower of massive states. There is a massless graviton and an infinite
set of massive spin 2 fields. In addition there are vectors and scalars.

The covariant derivatives ${\cal D}_\a$ are
\begin{equation}
\begin{split}
{\cal D}_\r g_{ab} & = \pa_\r g_{ab}- A^c_\r \pa_c g_{ab} -2 g_{c (a}  \pa_{b)}  A_\r ^c, \\
{\cal D}_\r h_{\m\n} & = \pa_\r h_{\m\n}-  A^c_\r  \pa_c  h_{\m\n} -2 k
h_{\m\n} \hat \nabla _c A^c_\r.
\end{split}
\end{equation}
Note that for simplicity we have used the same symbol $h_{\m\n}$ to denote two different fields. The field used here differs from the one used in sections 2.1 and 2.2 by a factor of $f$.

\section{Anti-symmetric tensor fields}

The standard kinetic term for antisymmetric tensors $F = dC$ in $D$ dimensions is
\begin{equation}
S_{\rm tensor}=
-\frac{1}{ 4 \k^2 }\int d^D x\sqrt{-G}{1\o p!} G^{M_1N_1}\cdots G^{M_p N_p}F_{M_1\cdots M_p}F_{N_1\cdots N_p}.
\end{equation}
We wish to construct the $d$-dimensional space-time effective action.

\subsection{Compactification}

The cleanest method to obtain the space-time effective action is to use the following basis of the
cotangent space
\begin{equation}
\begin{split}
Dy^\a&  = dx^\a, \\
Dy^a & = dy^a + A^a_\a dx^\a,
\end{split}
\end{equation}
and the dual basis of the tangent space
\begin{equation}
\begin{split}
D_\a &={D \o Dy^\a} =  {\pa \o \pa x^\a} - A_\a^a {\pa \o \pa y^a}, \\
D_a  &= {D \o D y^a} = {\pa \o \pa y^a}.
\end{split}
\end{equation}
We call this the ``new basis''.
Given the pairing $\langle\;,\; \rangle:T^\star_p Y \times T_p Y \to \RR $ with
\begin{equation}
\langle dy^N, {\pa \o \pa y^M}\rangle = \d^N_M,
\end{equation}
the new basis was chosen such that
\begin{equation}
\langle Dy^N, {D \o D y^M}\rangle = \d^N_M,
\end{equation}
for indices $M,N\dots=1,\dots,D$. We use the symbol $y^N$ to label the
coordinates of the $D$-dimensional space.  Our index conventions are
explained in footnote 1.

A differential $n$-form $\ome$ can then be expanded in either basis
\begin{equation}
\omega= {1\o n!} \omega_{N_1 \dots N_n} dy^{N_1} \wedge \dots \wedge dy^{N_n}
= {1\o n!} {\widetilde \ome}_{N_1 \dots N_n} Dy^{N_1} \wedge \dots \wedge Dy^{N_n}.
\end{equation}
Explicitly the components of a differential form with $r$ indices parallel to $Y$ and $s$ indices
parallel to $M^d$ are
\begin{equation}
{\widetilde \ome}_{a_1 \dots a_r \a_1 \dots \a_s} = \ome_{a_1 \dots a_r A_1 \dots A_s}
(\d_{\a_1}^{A_1} - A_{\a_1}^{b_1} \d_{b_1}^{A_1})\dots (\d_{\a_s}^{A_s} - A_{\a_s}^{b_s} \d_{b_s}^{A_s}).
\end{equation}
The inverse relation is
\begin{equation}
\ome_{a_1 \dots a_r \a_1 \dots \a_s} = {\widetilde \ome}_{a_1 \dots a_r A_1 \dots A_s}
(\d_{\a_1}^{A_1} + A_{\a_1}^{b_1} \d_{b_1}^{A_1})\dots (\d_{\a_s}^{A_s} + A_{\a_s}^{b_s} \d_{b_s}^{A_s})
\end{equation}
In order to avoid cluttering the equations we sometimes label the components of tensors in the new basis by
typewriter letters, for example $\tt C$, $\tt F$.
The components of the
exterior derivative of a differential $n$-form expanded in the new basis are
\begin{equation}
\begin{split} \label{dercom}
{\widetilde{( d \ome)}}_{N_1 \dots N_{n+1}} = & (n+1) D_{[N_1} {\widetilde \ome}_{N_2 \dots N_{n+1}]} \\
+ &{1\o 2}  n(n+1) {\cal F}^a_{\a\b} {\widetilde \ome}_{a [N_1 \dots } \d^\a_{N_n} \d^\b_{N_{n+1}]} \\
+ & n (n+1) \pa_a A^b_{\a } {\widetilde \ome}_{b [N_1 \dots } \d^a_{N_n} \d^\a_{N_{n+1}]}.
\end{split}
\end{equation}

As an illustrative example lets work out the details of a three-form potential. The components in the two bases are related by
\begin{equation}
\begin{split}
{\tt C}_{abc} & = C_{abc},\\
{\tt C}_{\m ab} & =C_{ab\m} - A_\m^c C_{abc}, \\
{\tt C}_{\m\n a} & = C_{a\m\n} - A_\m^b C_{ab\n } - A_\n^b C_{a\m b }+A^b_\m A^c_\n C_{abc},\\
{\tt C}_{\m\n\r} & =C_{\m\n\r} -3 A_{[\m}^a C_{\n\r ]a} +3 A^a_{[\m} A^b_{\n} C_{\r] ab}+A^a_{[\m} A^b_{\n} A^c_{\r]} C_{abc}.\\
\end{split}
\end{equation}

\subsection{Internal diffeomorphisms}

The transformation properties of the tensors ${\widetilde T}$
under infinitesimal coordinate transformations
$y'^a= y^a - \xi^a(x,y)$ are
\begin{equation}\label{tsss}
\d {\widetilde T}_{a_1 \dots a_p \a_1 \dots \a_q} = \xi^r \pa_r  {\widetilde T}_{a_1 \dots a_p \a_1 \dots \a_q} +
\sum_{k=1}^p \pa_{a_k} \xi^r {\widetilde T}_{a_1 \dots r \dots a_p \a_1 \dots \a_q}.
\end{equation}
In abbreviated notation we set
\begin{equation}
{\tt C}_{\m_1 \dots \m_n (a_1 \dots a_{p-n} ) } = {\tt C}_{\m_1 \dots \m_n a_1 \dots a_{p-n}}(x,y),
\end{equation}
and the infinitessimal change of ${\tt C}$ takes the form
\begin{equation}
\d  {\tt C}_{\m_1 \dots \m_n (b_1 \dots b_{p-n} ) }= \left( t^{(n)}_{a'}\right)_{(b_1 \dots b_{p-n})}^{
\hph{(b_1 \dots b_{p-n})} (c_1 \dots c_{p-n})'' }{\tt C}_{\m_1 \dots \m_n (c_1 \dots c_{p-n} )'' }\l^{a'},
\end{equation}
with
\begin{equation} \label{reps}
\begin{split}
& \lp t^{(n)}_{a'}\rp_{(b_1\cdots b_{p-n})}^{\hph{(b_1\cdots b_{p-n})'}
(c_1\cdots c_{p-n})''} =\d^{[c_1}_{[b_1}\cdots\d^{c_{p-n}]}_{b_{p-n}]}
\d(y-y')\pa_a\d(y-y'')\\
& +\lp -1\rp^{p-n+1}\lp p-n\rp\d^{[c_1}_a\d^{c_2}_{[b_1}\cdots\d^{c_{p-n}]}_{b_{p-n-1}}\pa_{b_{p-n}]}\d(y-y')\d(y-y'').
\end{split}
\end{equation}

So, for example, the infinitessimal change of the three-form is
\begin{equation} \label{dcsai}
\begin{split}
\d {\tt C}_{abc}& = \xi^r \pa_r {\tt C}_{abc} + 3 \pa_{[a} \xi^r {\tt C}_{bc] r}, \\
\d {\tt C}_{\a ab}& = \xi^r \pa_r {\tt C}_{\a ab} + 2 {\tt C}_{\a r[b} \pa_{a]} \xi^r,\\
\d {\tt C}_{\a \b a}& =\xi^r \pa_r {\tt C}_{\a\b a} + \pa_a \xi^r {\tt C}_{\a\b r} ,\\
\d {\tt C}_{\a \b \g}& =\xi^r \pa_r {\tt C}_{\a \b \g},\\
\end{split}
\end{equation}
after a coordinate transformation with parameter $\xi^r$. Using eqn.\ \C{covi} this can be used to define
derivatives ${\cal D}_\a$, which
transform covariantly under internal diffeomorphisms. We find
\begin{equation}
\begin{split}
{\cal D}_\a {\tt C}_{a b c  }  & =
\pa_\a {\tt C}_{abc}-A^d_\a \pa_d {\tt C}_{abc} -3 {\tt C}_{d[ab} \pa_{c]} A^d_\a , \\
{\cal D}_\a {\tt C}_{\m a b  }  & = \pa_\a {\tt C}_{\m a b}-A_\a^c \pa_c {\tt C}_{\m a b}+2
{\tt C}_{\m c [ a }\pa_{b]}   A_\a^c , \\
{\cal D}_\a {\tt C}_{\m\n a }  & = \pa_\a {\tt C}_{\m\n a }-A_\a^b \pa_b {\tt C}_{\m\n a}-
{\tt C}_{\m\n b}\pa_a A_\a^b , \\
{\cal D}_\a {\tt C}_{\m\n\r}  & = \pa_\a {\tt C}_{\m\n\r}-A_\a^a \pa_a {\tt C}_{\m\n\r} . \\
\end{split}
\end{equation}

\subsection{Gauge transformations}

The gauge invariance of the $p$-th rank antisymmetric tensor $\d C = d \L$ with $\L$ being a
$(p-1)$-form leads to invariances of the space-time effective action. So for example, if $p=3$ we need to consider 3 different transformations, by parameters
\begin{equation}\label{las}
\L_{ab}, \qquad \L_{\a a},\qquad {\rm and } \qquad \L_{\a\b}.
\end{equation}
To avoid cluttering the formulas we use the same symbol $\L$ for all transformations. The different types
of space-time transformations are specified by the index structure of $\L$.
We find
\begin{equation} \label{dcsai}
\begin{split}
\d {\tt C}_{abc}& = 3 \pa_{[a} {\widetilde \L}  _{bc]}, \\
\d {\tt C}_{\a ab}& =  {\cal D}_{\a} {\widetilde \L} _{ab}+ 2 \pa_{[a} {\widetilde \L}_{b]\a},\\
\d {\tt C}_{\a \b a}& =
2 {\cal D}_{[\a} {\widetilde \L}  _{\b]a }+\pa_a {\widetilde \L}_{\a\b}-{\widetilde \L}  _{ab}
{\cal F}_{\a\b}^b  ,\\
\d {\tt C}_{\a \b \g}& =3 {\cal D}_{[\a} {\widetilde \L}  _{\b \g ]}+ 3 {\widetilde \L}  _{a[\a} {\cal F}_{\b\g]} ,\\
\end{split}
\end{equation}
We note that the components of $\L$ in eqn.\ \C{las} are tensors in $D$ dimensions. So
the components of ${\widetilde \L}$ change according to eqn.\ \C{tsss} after an internal diffeomorphism.
Correspondingly the components of ${\widetilde \L}$ are in the representation \C{reps}.
We have defined the covariant derivatives of ${\widetilde \L}$ accordingly
\begin{equation}
\begin{split}
{\cal D}_{\m_1 }{\widetilde \L}_{\m_2 \dots \m_n  b_1 \dots b_{p-n} }  & =
\pa_{\m_1 }{\widetilde \L}_{\m_2 \dots \m_n  b_1 \dots b_{p-n} }
-A^a_{\m_1} \pa_a {\widetilde \L}_{\m_2 \dots \m_n b_1 \dots b_{p-n}} \\
& + (p-n) (-1)^{p-n}{\widetilde \L}_{\m_2 \dots \m_n a [b_1 \dots b_{p-n-1}}\pa_{b_{p-n}]} A_{\m_1}^a.
\end{split}
\end{equation}

We can write
\begin{equation}\label{ctraf}
\begin{split}
\d {\tt C}_{\m_1 \dots \m_n (b_1 \dots b_{p-n})}  & =
n {\cal D}_{[\m_1} {\widetilde \L}_{\m_2 \dots \m_n] (b_1 \dots b_{p-n})}\\
& + \left( q^{(n)} \right) _{(b_1 \dots b_{p-n} )} ^{\hph{(b_1 \dots b_{p-n} )}(c_1 \dots c_{p-n-1})'}
{\widetilde \L}_{\m_1 \dots \m_n (c_1 \dots c_{p-n-1})'} \\\o &
+ {n(n-1)\o 2} \left( h^{(n)} _{a'} \right) _{(b_1 \dots b_{p-n} )} ^{\hph{(b_1 \dots b_{p-n} )}(c_1 \dots c_{p-n+1})''}{\cal F}^{a'}_{[\m_1 \m_2} {\widetilde \L}_{\m_3 \dots \m_n] (c_1 \dots c_{p-n+1})''}.
\end{split}
\end{equation}
Comparing with eqn.\ \C{dcsai} we find
\begin{equation}
\begin{split}
\lp q^{(n)}\rp_{(b_1\cdots b_{p-n})}^
{\hph{(b_1\cdots b_{p-n})}(c_1\cdots c_{p-n-1})'}=
\lp -1\rp^{p-1}\lp p-n\rp\d^{[c_1}_{[b_1}\cdots\d^{c_{p-n-1}]}_{b_{p-n-1}}\pa_{b_{p-n}]}\d(y-y'),
\end{split}
\end{equation}
and
\begin{equation}
\begin{split}
\lp h^{(n)}_{a'}\rp_{(b_1\cdots b_{p-n})}^{\hph{(b_1\cdots b_{p-n})'}(c_1\cdots c_{p-n+1})''}=
\lp -1\rp^n\d^{[c_1}_a\d^{c_2}_{[b_1}\cdots\d^{c_{p-n+1}]}_{b_{p-n}]}\d(y-y')\d(y-y'').
\end{split}
\end{equation}
We have written these expressions in a form that applies to a $p$-th rank antisymmetric tensor
with any number of space-time indices. So to compare with eqn.\ \C{dcsai} set $p=3$ and take $n=0,\dots,3.$

The components of $F=d C$ expressed in the new basis are covariant under internal diffeomorphisms in the sense that these transform according to eqn.\ \C{tsss} and
are invariant under the gauge transformations of the antisymmetric tensor.
For $p=3$ we find the corresponding field strengths
\begin{equation}\label{fsai}
\begin{split}
{\tt F}_{abcd} & = 4 \pa_{[a} {\tt C}_{bcd]}, \\
{\tt F}_{\m a b c} & = {\cal D}_{\m} {\tt C}_{abc} -3 \pa_{[a} {\tt C}_{bc] \m}, \\
{\tt F}_{\m\n  a b} & = 2 {\cal D}_{[\m } {\tt C}_{\n ] a b}
+ 2\pa_{[a} {\tt C}_{b] \m\n }+{\cal F}_{\m\n }^c {\tt C}_{abc}, \\
{\tt F}_{\m\n \r  a} & =3 {\cal D}_{[\m} {\tt C}_{\n \r ] a}- \pa_a {\tt C}_{\m\n \r }
+ 3 {\cal F}^b _{[\m \n  }{\tt C}_{\r] ab }  , \\
{\tt F}_{\m\n \r \s } & = 4 {\cal D}_{[\m } {\tt C}_{\n \r \s ]} + 6 {\cal F}^a _{[\m\n } {\tt C}_{\r \s ] a}.
\end{split}
\end{equation}
We note $\pa_a = D_a$ and moreover, covariant derivatives $\hat \nabla_a$ could have been used instead of $\pa_a$ since the connections drop out when anti-symmetrizing.
And in general
\begin{equation}
\begin{split}
{\tt F}_{\m_1\cdots\m_{n+1}(b_1 \dots b_{p-n})}& =
\lp n+1\rp{\cal D} _{[\m_1}{\tt C}_{\m_2\cdots\m_{n+1}](b_1 \dots b_{p-n}) }\\
& -\lp q^{(n)} \rp_{(b_1 \dots b_{p-n})} ^{\hph{(b_1 \dots b_{p-n})}(c_1 \dots c_{p-n-1})' }
{\tt C}_{\m_1\cdots\m_{n+1}(c_1 \dots c_{p-n-1})'}\\
& -\frac{n(n+1)}{2}
\lp h^{(n)}_{a'}\rp_{(b_1 \dots b_{p-n})}^{\hph{(b_1 \dots b_{p-n})}(c_1 \dots c_{p-n+1})''}
{\cal F}^{a'}_{[\m_1\m_2}{\tt C}_{\m_3\cdots\m_{n+1}](c_1 \dots c_{p-n+1})''}.
\end{split}
\end{equation}

\subsection{The kinetic and potential terms}

Next we present the part of the effective space-time action arising from the kinetic term
of a $p$-th rank antisymmetric tensor in $D$ dimensions. We first present our result for eleven-dimensional supergravity, elaborating further the $p=3$ example, and then we present the result for general $p$. As an illustrative example we also choose the space-time dimension $d=4$.

The form action in eleven-dimensional supergravity is
\begin{equation}
S= -{1\o 4 \k^2} \int d^{11} x \sqrt{-G} \mid F \mid^2.
\end{equation}
The eleven-dimensional three-form $C$ gives rise to scalars $C_{abc}$,
vectors $C_{a b \m}$ and tensors $C_{ a \m \n }$ and $C_{\m\n\l}$ in $d$ dimensions. In the following we explain the choice of space-time fields and we present the effective action. We find
\begin{equation}
S = S_{\rm s} + S_{\rm v} + S_{\rm t},
\end{equation}
{\it i.e.} the action is the sum of the contribution from scalars, vectors and tensors.

First, there are space-time
scalars ${\tt C}_{abc}$ which are three-forms on $Y$. The space-time action is\footnote{We are using another shorthand notation here. We use the notation $(\dots)^2$ whenever the indices are contracted
in an obvious way with $h^{\m\n}$ and/or $g^{ab}$.
Explicitly, given any quantity with some space-time and some internal indices we identify
$(W_{\m \n  \dots a b \dots})^2$ with $ W_{\m_1 \n_1  \dots a_1 b_1 \dots} W_{\m_2 \n_2  \dots a_2 b_2 \dots}h^{\m_1 \m_2} h^{\n_1 \n_2} \dots g^{a_1 a_2} g^{b_1 b_2}\dots . $}
\begin{equation}\label{acs}
\begin{split}
S_{\rm s}& =   - {1\o 24 \k^2} \int d v \left[
\left( {\cal D}_{\m} {\tt C}_{abc} -3 \pa_{[a} {\tt C}_{bc] \m}\right)^2
 +  4f \left(\hat{\nabla}_{[a}{\tt{C}}_{bcd]}\right)^2\right],
\end{split}
\end{equation}
where $f$ was defined in eqn.\ \C{eq:fDef}. The first bracket is the kinetic term for space-time scalars ${\tt C}_{abc}$ which are charged under
the non-abelian gauge group arising from internal diffeomorphisms $A_\m^a$
and the abelian gauge field ${\tt C}_{ab\m}$. The second term is a potential
for the scalars ${\tt C}_{abc}$ arising from the antisymmetric tensor
and the scalars arising from the internal metric $g_{ab}$.

We note that the variation of the last term in
${\tt F}_{\a\b a b}$ in eqn.\ \C{fsai} under \C{dcsai} does not involve a space-time derivative. We therefore find it more convenient to define a new gauge field tensor
\begin{equation}
{\cal F}_{\m\n a b} =  2 {\cal D}_{[\m } {\tt C}_{\n ] a b}
+ 2\pa_{[a} {\tt C}_{b] \m\n } .
\end{equation}
After this redefinition the gauge kinetic term is
\begin{equation}
S_{\rm v} =   - {1\o 16 \k^2} \int d v
f^{-1} \left( {\cal  F}_{\m\n a b}+ {\cal F}_{\m\n}^c {\tt C}_{abc} \right)^2.
\end{equation}
Moreover, the action for tensors is
\begin{equation}
S_{\rm t}  =   - {1\o 24 \k^2} \int d v \Big[
f^{-2}  \left( {\tt F}_{\m \n \r a } \right)^2  +
{1\o 4}f^{-3}\left(  {\tt F}_{\m\n\r\s} \right)^2.
 \Big]
\end{equation}

In general, the space-time effective action arising from a rank $p$
antisymmetric tensor in $D$ dimensions is
again the sum of the contribution of scalars, vectors and tensors
\begin{equation}
S=S_{\rm s} + S_{\rm v} + S_{\rm t}  .
\end{equation}
Using as a starting point the $D$-dimensional action
\begin{equation}\label{fgen}
S= \int d^D x \sqrt{-G} \mid F\mid^2,
\end{equation}
where $F=d C$ and $C$ is a $p$-form, we obtain\footnote{Here we work with the metric reduction \C{bi1}, without the warp factor $f$ needed to obtain canonical Einstein-Hilbert terms.  If we wanted to include these extra factors, as we did in the M-theory example above, then the measure factors $dV$ (defined in \C{eq:dV}) would need to be modified in a straight-forward way.}
\begin{equation}
\begin{split}
S_{\rm s}  = & {1\o p!} \int d V \left[ \left({\cal D}_\m {\tt C}_{a_1 \dots a_p} + p (-1)^{p} \pa_{[a_1}
 {\tt C}_{a_2 \dots a_p ] \m } \right)^2  +(p+1)\left(\hat{\nabla}_{[a_1}{\tt{C}}_{a_2\cdots a_{p+1}]}\right)^2\right]\\
S_{\rm v} = & {1\o 2(p-1)!} \int d V \left({\cal F}_{\m\n a_1 \dots a_{p-1} } +{\cal F}_{\m\n} ^b {\tt C}_{b a_1 \dots a_{p-1} }\right)^2   \\
S_{\rm t} = &  {1\o (p+1)!}\sum_{k=0}^{p-2} \begin{pmatrix} p+1 \\ k \end{pmatrix}\int dV
({\tt F}_{a_1 \dots a_k \m_{k+1} \dots \m_{p+1}})^2,\\
\end{split}
\end{equation}
Here using the same reasoning
as before we defined the gauge field ${\cal F}_{\m\n a_1 \dots a_{p-1}}$ according to
\begin{equation}
{\cal F}_{\m\n a_1 \dots a_{p-1}} = 2  {\cal D}_{[\m} {\tt C}_{\n]a_1 \dots a_{p-1} }+(p-1)\pa_{[a_1}
{\tt C}_{b_2 \dots b_{p-1}] \m\n}  .
\end{equation}
It is then easy enough to modify these concrete expressions to include overall scalar functions or numerical coefficients in eqn.\ \C{fgen}.

\subsection{Chern-Simons terms}

Supergravity and string effective actions also include in general Chern-Simons terms.
For example, in eleven-dimensional supergravity
\begin{equation}
S_{\rm cs} = -{1\o 12 \k^2} \int C \wedge F \wedge F.
\end{equation}
Let us first dimensionally reduce this action to four space-time dimensions.
The most practical method is to expand the differential forms in the new basis. Given,
for example, a vector $V$
in $D$ dimensions the relevant expansion is
\begin{equation}
V = V_N dy^N = (V_\a - A_\a ^a V_a) dx^\a + V_a (dy^a + A_\a^a dx^\a) = {\tt V}_\a dx^\a + V_a Dy^a.
\end{equation}
Applying this expansion to each index independently organizes the Chern-Simons term into
\begin{equation}
S_{\rm cs} = - {55 \o 2^8 3^3 \k^2} \int dx^{\m\n\r\s} dy^{abcdefg} {\tt F}_{[\m\n\r\s }
{\tt F}_{abcd} {\tt C}_{efg]} ,
\end{equation}
where $dx^{\m\n\dots} = dx^\mu \wedge dx^\n\wedge \dots$. The components of $F$ in the new basis can be found in eqn.\ \C{fsai} and the anti-symmetrization
is done over all indices, external and internal.  Here we have used
\begin{equation}
dx^{\m\n\r\s} Dy^{abcdefg} = dx^{\m\n\r\s} dy^{abcdefg} .
\end{equation}

For generic supergravity theories the Chern-Simons terms are contributions to the action,
which depend on some $p$-form potential $C$, yet are gauge invariant.
This potential can, for example, be a RR, NS-NS potential
or the three-form of eleven-dimensional supergravity. Schematically the Chern-Simons terms take the form
\begin{equation}
S_{\rm cs} \sim \int  C \wedge \Omega,
\end{equation}
where $\Omega$ is a closed form constructed from field strengths. For eleven-dimensional supergravity $\Omega = F \wedge F$, for example. Even though the concrete expressions depend on the $D$-dimensional theory that is being reduced and the space-time dimension $d$, the method described above applies in general. Since using this method it is straightforward to work out the contribution to the space-time effective action but the results are case dependent we only present concrete results for the eleven to four reduction discussed above.

\subsection{St\"uckelberg mechanism}

Typically, one would like to understand how to fix as much of our gauge freedom as possible and determine the physical spectrum, especially the space of massless fields.  In this section we only need to work to linearized order in the transformations and field strengths.  Let's start by considering the case $(D,d,p)=(11,4,3)$.  The variations are given in \C{dcsai} and the field strengths in \C{fsai}.  Take the space-time coordinate $x$ to be arbitrary but fixed.  According to Hodge's theorem the three-form $C_{abc}$ can be decomposed into harmonic, exact and co-exact pieces, and we can use the gauge freedom from $\Lambda_{ab}$ to gauge away the exact piece.  Once this is done, the kinetic terms ${\tt{F}}_{\m abc}^2$ for the exact three-forms become mass terms for the vectors ${\tt{C}}_{\m ab}$ which are non-closed two-forms on the internal space.  This is the usual St\"uckelberg mechanism in which the non-closed two-form ${\tt{C}}_{\m ab}$ becomes massive after ``eating" the exact piece of ${\tt{C}}_{abc}$.  In the same way a non-closed one-form ${\tt{C}}_{a\m\n}$ becomes massive after ``eating'' the exact piece of ${\tt{C}}_{ab\m}$ (equivalently the exact pieces of ${\tt{C}}$ can always be gauged away). The potential in eqn.\ \C{acs} gives a mass to any non-closed scalars arising from the three-form.  Thus at each level harmonic forms are all that remain as massless fields in four-dimensions; the remaining fields can either be gauged away or become massive.

More generally, we have the terms with coefficients $q^{(n)}$ in eqn.\ \C{ctraf}.  Each $q^{(n)}$ is a linear operator from the space of fields that are $(n+1)$-forms in the $d$-dimensional space-time to the space of fields which are space-time $n$-forms.  In our case these are the spaces of $(p-n-1)$-forms and $(p-n)$-forms, respectively, on the internal manifold, and $q^{(n)}$ is simply the usual exterior derivative.  As before, we can gauge away each field in the image of $q^{(n)}$, {\it i.e.}\ each exact form, and each field that is not in the kernel of $q^{(n)}$, {\it i.e.}\ non-closed forms, gets a mass via the St\"uckelberg mechanism.

The remaining massless fields lie in the space of closed forms modulo exact forms, {\it i.e.}\ in the de Rham cohomology groups $H^{p-n}$.  For a compact manifold $Y$, although the space of $(p-n)$-forms is infinite dimensional, the $H^{p-n}(Y)$ are finite dimensional, so we see that there are always a finite number of massless fields coming from the reduction of the $D$-dimensional $p$-forms.

\section{Summary}

As a summary we present a concrete example.
The space-time effective action for eleven-dimensional supergravity compactified to four dimensions is
\begin{equation}
\begin{split}
S =  & -   {1\o 8 \k^2} \int dv h ^{\a\b}
\left({1\o 2} g^{ab}g^{cd}+g^{ac}g^{bd}\right){\cal D}_\a g _{ab}{\cal D}_\b g_{cd}  \\
 & +   {1\o 2 \k^2} \int d v \left(
h ^{ \b \m }h ^{\g[ \r} h ^{\a] \n} - {1\o 2} h ^{\a\m} h ^{\b[\n} h ^{\g] \r}\right)
{\cal D}_\a h _{\b\g} {\cal D}_{\m} h _{\n \r }\\
&  +  {1\o 4 \k^2} \int d v f
 \Big[ g^{ab}h^{\a[\b}h^{\m]\n}\hat  \nabla_a {h }_{\a\b}\hat \nabla_b {h }_{\m\n}
-h^{\a\b}\lp {1\o 2}  g^{ab}g^{cd}+g^{ac}g^{bd}\rp
\hat \nabla_a h_{\a\b}\hat\nabla_b g_{cd}
 \\
 & +\left( g^{pt} g^{qu} g^{rs} -
{1\o 2} g^{ps} g^{qt} g^{ru} +
  g^{pr} g^{qu} g^{st}  \right)
\hat \nabla_r g _{pq} \hat \nabla_u g_{st}
\Big]-   {1\o 8 \k^2} \int dv  f^{-1}
\left( {\cal F}_{\m\n}^a\right) ^2 \\
&   - {1\o 24 \k^2} \int d v \left[
\left( {\cal D}_{\m} {\tt C}_{abc} -3 \pa_{[a} {\tt C}_{bc] \m}\right)^2
 + 4 f \left(\hat{\nabla}_{[a}{\tt{C}}_{bcd]}\right)^2\right]\\
&   - {1\o 16 \k^2} \int d v
f^{-1} \left( {\cal  F}_{\m\n a b}+ {\cal F}_{\m\n}^c {\tt C}_{abc} \right)^2
 - {1\o 24 \k^2} \int d v \Big[
f^{-2}  \left( {\tt F}_{\m \n \r a } \right)^2  +
{f^{-3}\o 4}\left(  {\tt F}_{\m\n\r\s} \right)^2
 \Big] \\
 &  - {55 \o 2^8 3^3 \k^2} \int dx^{\m\n\r\s} dy^{abcdefg} {\tt F}_{[\m\n\r\s }
{\tt F}_{abcd} {\tt C}_{efg]} .
\end{split}
\end{equation}
This is the action for the bosonic fields in all representations of the four-dimensional
Lorentz-group and all masses.
The components of $F$ in the new basis are summarized in eqn.\ \C{fsai}.
Given the expressions we presented it is easy to write then down
the space-time effective action for any values of $(D,d,p)$.

\section*{Acknowledgement}\addcontentsline{toc}{section}{Acknowledgement}

This work
was supported by the grants PHY-1214344 and NSF Focused Research Grant DMS-1159404 and by
the George P. and Cynthia W. Mitchell Institute for Fundamental Physics and Astronomy. We thank Edward Witten for discussions.

\newpage

\providecommand{\href}[2]{#2}\begingroup\raggedright
\endgroup
\end{document}